\documentclass[%
apl
 amsmath,amssymb,
 reprint,%
]{revtex4-1}

\usepackage[version=3]{mhchem} % Formula subscripts using \ce{}
\usepackage[T1]{fontenc}       % Use modern font encodings
\usepackage{graphicx}
\usepackage{float}
\usepackage{xspace}
\usepackage{amsmath}

\usepackage{verbatim}
\usepackage[table,xcdraw]{xcolor}

\usepackage{times}
\usepackage{ulem}

\usepackage{lineno,hyperref}
\modulolinenumbers[5]

\newcommand{\be}{\begin{equation}}
\newcommand{\ee}{\end{equation}}
\newcommand{\ba}{\begin{eqnarray}}
\newcommand{\ea}{\end{eqnarray}}

\def \ctw {$^{12}$C\xspace}
\def \cth {$^{13}$C\xspace}

\begin{document}

\title{Thermal conductivity of graphene isotope superlattices}

\author{Eric Whiteway}
\author{Michael Hilke}
\affiliation{Department of Physics, McGill University, Montr\'eal, Canada H3A 2T8}

\begin{abstract}
Graphene has a high intrinsic thermal conductivity and a high electron mobility. The thermal conductivity of graphene can be significantly reduced when different carbon isotopes are mixed, which can enhance the performance of thermoelectric devices. Here we compare the thermal conductivities of isotopic \ctw/\cth random mixes with isotope superlattices with periods ranging from 46 to 225 nm. Raman Opto-Thermal conductivity measurements of these superlattice structures show an approximately 50$\%$ reduction in thermal conductivity compared to pristine \ctw graphene. This average reduction is similar to the random isotope mix. The reduction of the thermal conductivity in the superlattice is well described by a model of pristine graphene and an additional quasi-one dimensional periodic interfacial thermal resistance of $(2.5\pm 0.5) \mathrm{\times 10^{-11} \,m^2 K/W}$ for the \ctw/\cth boundary. This is consistent with a large anisotropic thermal conductivity in the superlattice, where the thermal conductivity depends on the orientation of the \ctw/\cth boundary.
\end{abstract}

%\end{frontmatter}
\maketitle

%\linenumbers
\section{Introduction}

The novel electronic properties of graphene have produced a great deal of interest in the material for a number of applications\cite{novo05,geim09,coop12}. Experimental results have indicated high thermal conductivity for graphene\cite{chen12,bala08,ghos08} which could make it an important material for heat management in electronic devices. In areas such as solid state refrigeration and thermoelectric power generation it's desirable to have materials with a combination of high electrical conductivity and reduced thermal conductivity\cite{nika17}. Isotope doping represents a powerful way to reduce thermal conductivity of materials without impacting their electronic properties.

Isotopically modified graphene has been shown to have reduced thermal conductivity relative to natural and isotopically pure graphene\cite{chen12} and has also been shown to increase optical phonons scattering\cite{rodr12}. The high Seebeck coefficient reported for graphene nanoribbons\cite{drag07} and high carrier mobility of graphene\cite{novo04,bolo08}, when combined with reduced thermal conductivity imposed by isotope impurities for  could lead to a unique high performance material for thermoelectric devices \cite{ouya09}. Isotope labelling\cite{li092,li10,li13,whit17} has proven to be a useful tool in tracking graphene CVD growth and demonstrates the ability to controllably introduce isotope doping at small time and spatial length scales

These isotope impurities also reduce thermal conductivity through the mass difference phonon scattering of individual atoms. In general, the phonon contribution to the thermal conductivity is given by  $K=\frac{1}{2}Cv\lambda$, where $C$ is the specific heat capacity, $v$ is the phonon group velocity and $\lambda$ the phonon mean free path (mfp) \cite{nika12}. Substitution of isotope atoms leads to an increase in the phonon point-defect scattering rate and a corresponding reduction in phonon mfp. In the case of \ctw/\cth graphene and when considering only isotope substitution and neglecting coherence effects, we can express the inverse mfp as $\lambda^{-1}= \rho^{12}\lambda_{12}^{-1}+\rho^{13}\lambda_{13}^{-1}+\Gamma$, where \cite{chen12}
\begin{equation}
\Gamma\sim \rho^{12}(1-M^{12}/\overline{M})^2+\rho^{13}(1-M^{13}/\overline{M})^2.
\end{equation}
$\lambda_{12}\simeq\lambda_0$ and $\lambda_{13}\simeq\lambda_0$ of the respective pure isotope lattices are similar. The same is true for their respective group velocities and specific heat. Hence, the dependence of the thermal conductivity on isotopes is dominated by the term $\Gamma$. 
$\Gamma\sim\rho^{12}(1-M^{12}/\overline{M})^2+\rho^{13}(1-M^{12}/\overline{M})^2$ which leads to a maximum scattering rate at approximately $50\%$ concentration. 

\begin{figure}[htbp]
\includegraphics[width=3.2in]{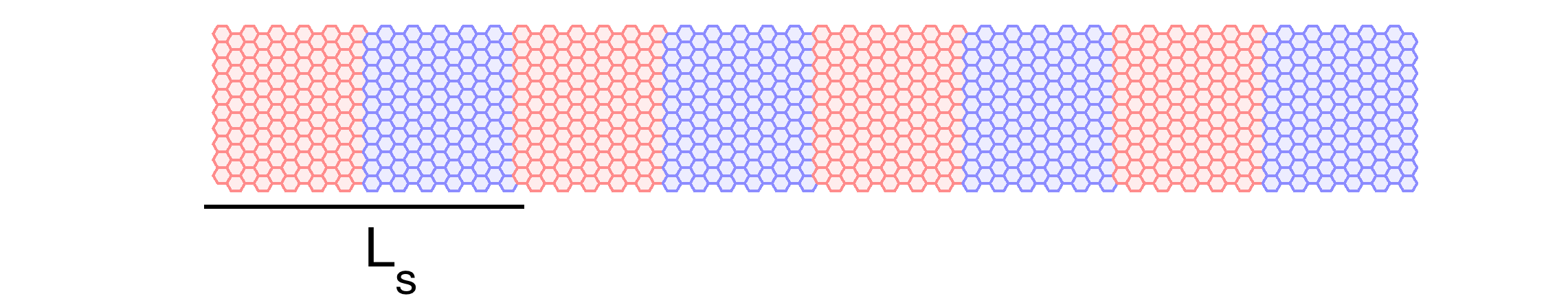}
\includegraphics[width=3.2in]{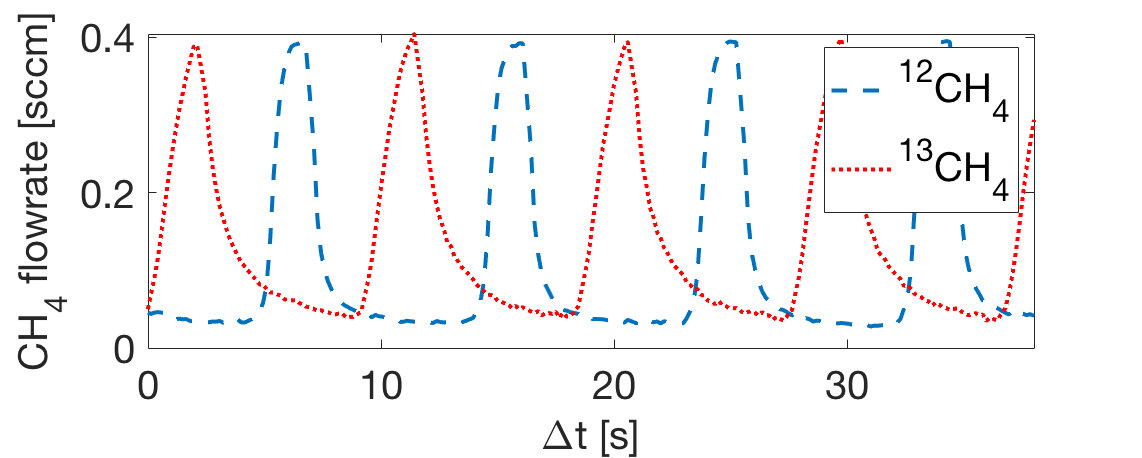}
\includegraphics[width=3.2in]{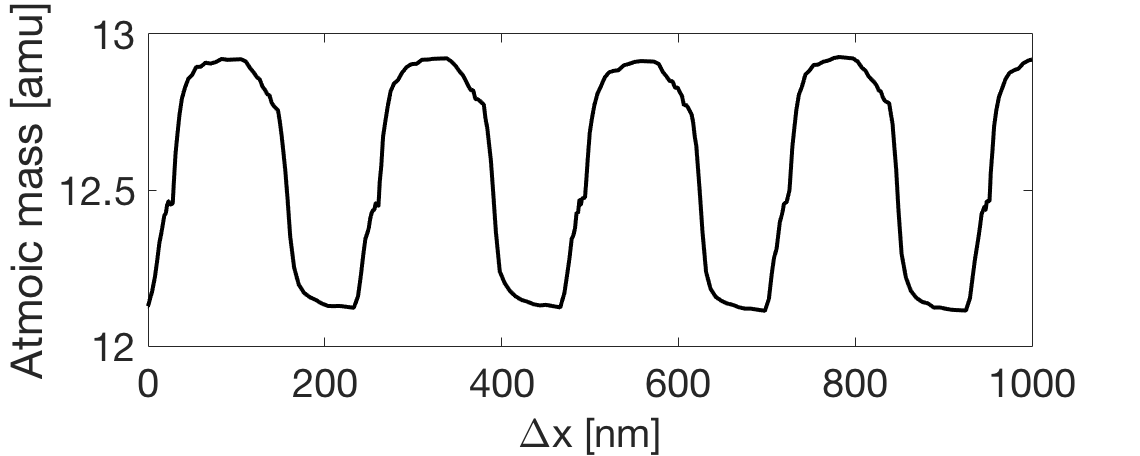}
\caption{Top: illustration of a graphene isotope superlattice with period $L_s$ of 5nm, where the blue regions correspond to \ctw and the red regions to \cth. Growth log showing flowrates $\dot{V}$ of \ctw and \cth methane along with the corresponding atomic mass vs. distance where distance is calculated as $\dot{V}\Delta t$ and scaled to correspond to the measured growth rate and atomic mass is measured as a function of isotopic methane concentration.}
\label{gasflows}
\end{figure}

The situation for ordered isotope superlattices (SLs) illustrated in figure \ref{gasflows} is different, since there are no random impurities. However, isotopic SLs are expected to have reduced thermal conductivities relative to isotopically pure materials \cite{sim00} and were shown to have a dependence on the superlattice period, with a minimum thermal conductivity corresponding to the crossover between coherent and incoherent phonon transport\cite{mu15}. At large periods, exceeding the phonon coherence length, the SL acts as a series of independent barriers, characterized by an interface density,  $I_d=\frac{1}{L_s}$ orthogonal to the periodicity. The thermal resistance is then expected to be proportional to $\sim I_dR_I$, where $R_I$ is the interfacial thermal resistance (Kapitza resistance) at the \ctw/\cth boundary\cite{mu15}. This breaks down when the S period is smaller than the phonon coherence length,  In the coherent regime the SL can no longer be seen as independent scatterers, but rather as a hybridized supercell, which leads to an expected increase in thermal conductivity relative to the superlattice minimum \cite{sim00}.

There have been several experimental realizations of SLs in three dimensional systems, such as Si/Ge\cite{borc00}, perovskite layers composed of $SrTiO_3$, $CaTiO_3$ and $BaTiO_3$ \cite{ravi13} and \ctw/\cth diamond\cite{wata11}. In these experiments, a minimum in thermal conductivity as a function of interface density was observed as well as  a reduced thermal conductivity relative to their bulk constituents or alloys of the two materials\cite{lee97,venk00}.

Graphene isotopic SLs have been studied theoretically using molecular dynamics (MD) \cite{hu10,mu15,davi17} and non-equilibrium Green's function methods\cite{ouya09}. The thermal conductivity of graphene isotope SL structure was found to be smaller relative to pristine graphene, with a minimum in thermal conductivity at the crossover between coherent and incoherent phonon transport occurring at a SL period of 6.25 nm \cite{mu15}. MD simulations have found that the reduction in thermal conductivity can be even larger for an isotope SL than for random isotope impurities \cite{hu10}.

The presence of thermal interface resistance results in a discrete temperature drop across the interface as has been observed in non-equilibrium molecular dynamics simulations of graphene isotope junctions \cite{pei12,mu15}. The calculated value of the thermal resistance of the \ctw/\cth  interface was found to be $1.05\times10^{-11}\,\mathrm{m^2 K/W}$\cite{pei12} and $3.88\times10^{-11}\,\mathrm{m^2 K/W}$ \cite{mu15}. The authors also found a dependence on the direction, {\it i.e.} \ctw to \cth vs. \cth to \ctw. Similar results were found for systems composed of graphene-hBN SLs\cite{feli18} and \ctw-$^{24}C$ nanoribbons\cite{xie17}. With additional isotope impurities on top of the SL structure the thermal conductivity was shown to be further reduced \cite{gu18}

Here we report the thermal conductivity measurements of graphene monocrystals synthesized by chemical vapor deposition with an artificial isotope SL with periods $L_s$ ranging from 46 to 225 nm. Because of the much smaller phonon coherence length than $L_s$, we expect the SL to be in the incoherent phonon transport regime \cite{mu15}, which is dominated by the added interfacial thermal resistance at each half period. We therefore expect the thermal conductivity to monotonically  decrease with increased interface density, $I_d$.

\section{Synthesis and Raman spectroscopy}

Superlattice synthesis was accomplished by chemical vapour deposition, alternatively pulsing \ctw and \cth methane gases on $\sim$1 second timescales. Graphene CVD substrate was a commercially available 25 $\mu$m thick copper foil and gas stock consisted of $^{12}$C methane (99.99\% purity) or $^{13}$C-methane (99.9\% purity) from Cambridge Isotopes Laboratories (CLM-3590-1). The resulting graphene single crystals were transferred using a typical PMMA wet transfer technique to either Si/SiO$_2$ wafers or holey SiN substrates for analysis by Raman spectroscopy.

Results are based on a single sample where the methane isotope dosing sequence was varied as a function of time. The sample is divided into 6 distinct regions corresponding to, pure \ctw, 50$\%$ \ctw -\cth mix, and 4 SLs of varying periodicity. The regions are separated by small regions of pure \ctw or \cth graphene in order identify the corresponding dosing sequence and extract the SL period $L_s$. Figure \ref{gasflows} shows a typical gas flow sequence along with the associated isotope distribution as a function of radial distance for the SL region.

\begin{figure}[htbp]

\includegraphics[width=.9\columnwidth]{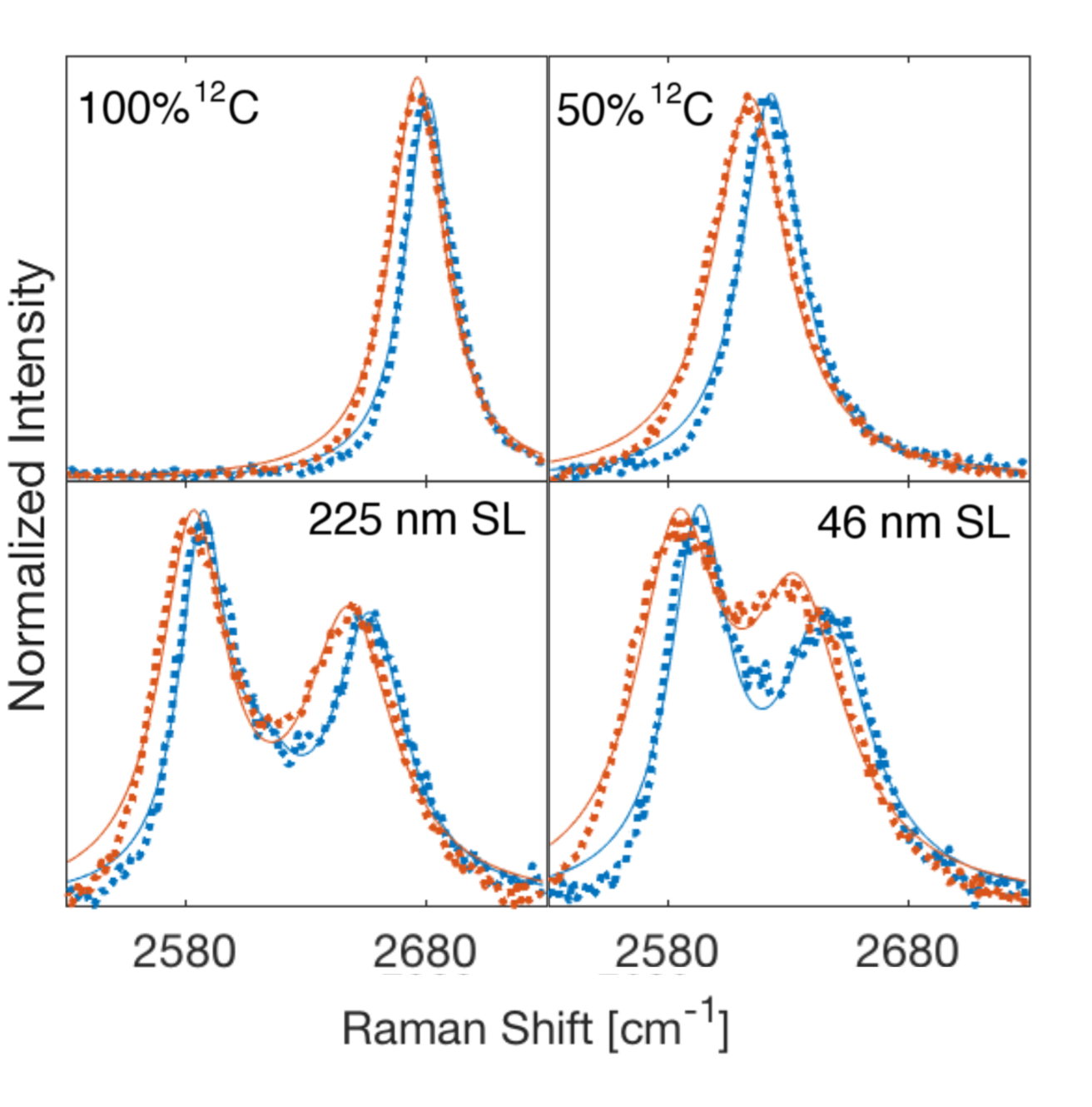}
\caption{2D peak Raman spectra for high (red) and low (blue) power laser excitations. The dots are the experimental spectra, while the lines correspond to Lorentzian fits. The top left is for pure \ctw, top right for the homogenous \ctw-\cth mix, bottom left for the 225nm period, and bottom right the 46nm period. The magnitude of the shift in peak position (red vs blue) is proportional to the graphene membrane temperature.}
\label{spectra}
\end{figure}

\begin{figure}[ht]
\begin{center}
\includegraphics[width=.99\columnwidth]{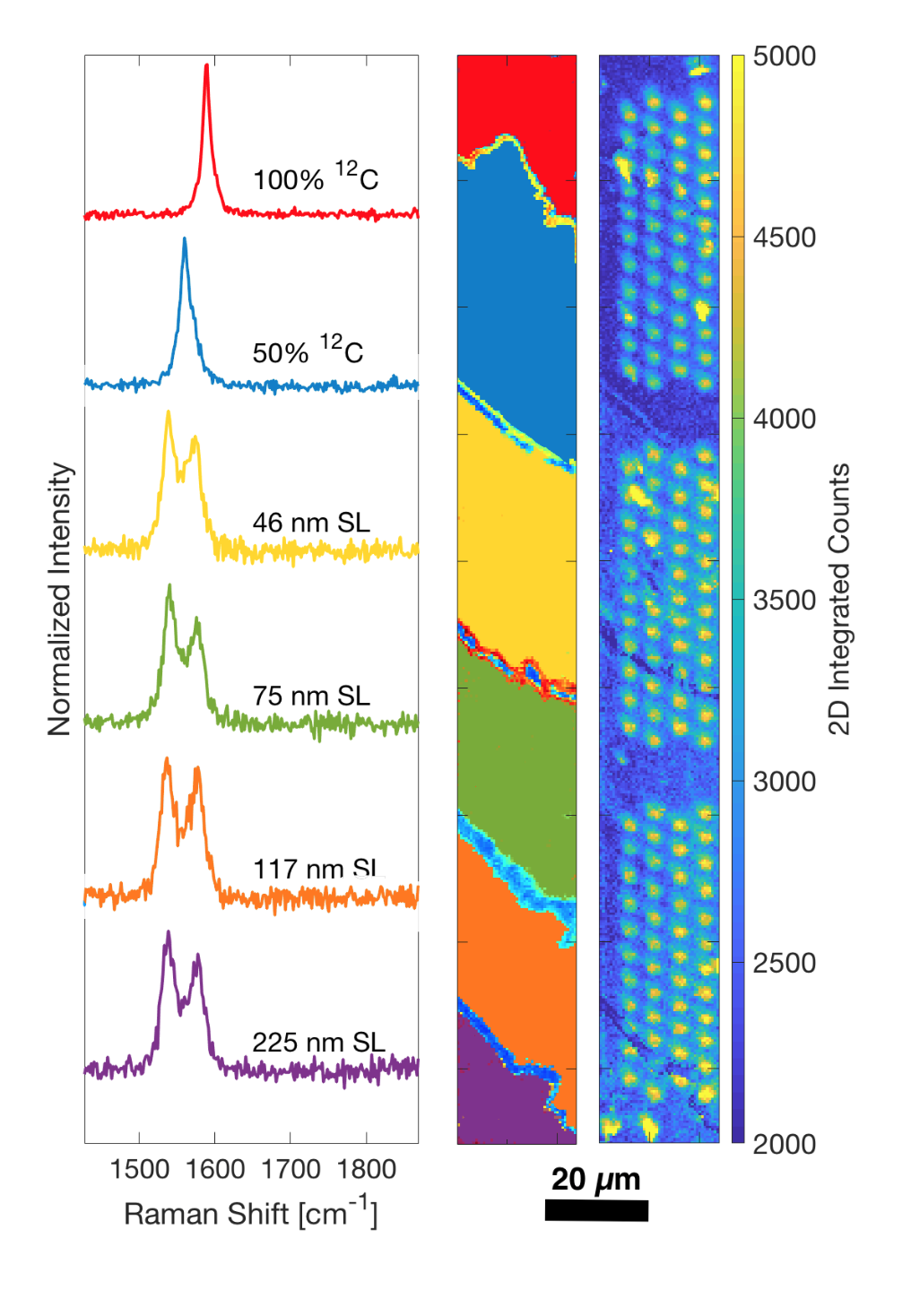}

\caption{a) Raman G peak for 6 distinct regions of Isotope superlattice sample. b) Corresponding regions c) Integrated 2D peak counts Raman maps of graphene on holey membrane. Suspended graphene shows an increase in 2D peak intensity consistent with previous results showing increase 2D peak intensity for suspended vs supported graphene\cite{ni09}.}
\label{optofig}
\end{center}

\end{figure}

Samples were characterized by Raman spectroscopy (figure \ref{spectra}) using a Renishaw Invia system and a 514 nm laser excitation source. The SL period, $L_s$ is extracted from the Raman map (figure \ref{optofig}) as a function of the length of a given region and the isotope dosing sequence as characterized in figure \ref{gasflows}.

The Raman spectra of different regions of the sample reflect their isotope distribution (see figure \ref{optofig}). For the pure \ctw region we observe a single narrow lorentzian G or 2D-peak, whereas for the 50$\%$ we observe significant broadening consistent with an increase in phonon scattering\cite{rodr12} (see figures \ref{spectra} and \ref{optofig}). In the superlattice regions we observe a double peak structure which is roughly the sum of the bulk \ctw and \cth Raman sectra. This indicates the formation of a heterogenous isotope distribution with a periodic variation in the phonon local density of states. This also shows that the optical phonon coherence length is smaller than the superlattice period, otherwise hybridization of the phonon bands would occur, which would narrow the separation between the heterogenous Raman peaks. 
=

\section{Raman Opto-thermal Conductivity measurement}

Graphene superlattices were suspended on gold covered holey SiN membranes and thermal conductivity measured using the Raman opto-thermal technique as previously described\cite{bala08,ghos08,cai10,chen12,li17}. This technique uses the Raman excitation laser to heat the suspended graphene sheet. The temperature dependance of the Raman 2D peak shift is used to measure the local temperature rise in the suspended film and the dependance of this temperature rise on absorbed laser power is used to extract the thermal conductivity.  The details of the thermal conductivity measurement were based off the approach of Cai et al\cite{cai10} and Chen et. al\cite{chen12} who used a similar arrangement of Au covered holey SiN membranes to measure the thermal conductivity of various isotopic mixes. The 2D peak temperature dependence is taken as $-7.23 \times 10^{-2}\,cm^{-1}/K$ for \ctw graphene and $-6.98 \times 10^{-2} \,cm^{-1}/K$ for the 50\% mix and superlattice samples\cite{chen12}. We similarly use a value for laser absorption of 3.4$\%$ and a laser spot size of 340 nm. The thermal conductivity is then obtained as:
\begin{equation}
K= \alpha \frac{ln \left( \frac{R}{r_0} \right)}{2\pi t R_g}
\label{k}
\end{equation}

Where $R= 1\,\mu$m is the radius of the hole, $r_0=170$ nm is the radius of the laser spot, $t=3.4$ \AA\xspace is the thickness of the graphene film and $R_g$ is the measured thermal resistance. $\alpha=0.96$ is a constant that is a function of $R$ and $r_0$\cite{cai10}.

\begin{equation}
R_g = \frac{T_m - T_a}{Q_{abs}}
\label{rg}
\end{equation}
Where $T_m$ is the measured temperature of the film and $T_a$ is the ambient temperature. We neglect heat loss to the environment and thermal contact resistance between the graphene and the gold substrate.

In the case of pure \ctw graphene we obtain a value of $K=3200\pm 1200$ W/m-K. We observe a reduction in the thermal conductivity for both the homogenous isotope mixtures of 1800$\pm$600 W/m-K for 50\% \ctw and the periodic superlattices with $L_s$ from 46-225 nm where we find $K$ between 1700 and 2100 W/m-K. These values correspond to the lowest temperature measurement and cover a range from approximately 316-335 K. These values are consistent with previous reports for both 100\% and 50\% \ctw which were reported as 4120 and 1977 W/m-K respectively\cite{chen12} at similar temperatures. The Thermal conductivity values as a function of heating power are tabulated in table \ref{reftable}.

\begin{table*}[htb]

\begin{tabular}{r|cc|cc|cc|cc|cc|cc|c}

& \multicolumn{2}{c|}{\textbf{100\% \ctw}} &\multicolumn{2}{c|}{\textbf{50\% \ctw}}&\multicolumn{2}{c|}{\textbf{46 nm SL}}&\multicolumn{2}{c|}{\textbf{75 nm SL}}&\multicolumn{2}{c|}{\textbf{117 nm SL}}&\multicolumn{2}{c|}{\textbf{225 nm SL}}& $\mathbf{R_{int}}$\\

Q [mW]&\textit{T[K]}&{K [W/m-K]}&\textit{T[K]}&{K [W/m-K]}&\textit{T[K]}&{K [W/m-K]}&\textit{T[K]}&{K [W/m-K]}&\textit{T[K]}&{K [W/m-K]}&\textit{T[K]}&{K [W/m-K]}& [$\mathrm{m^2 K/W}$]\\\hline
\rowcolor[HTML]{EFEFEF}[\tabcolsep] 
{2.50}&\textit{316}&{3242}&\textit{332}&1793&\textit{335}&1703&\textit{333}&1770&\textit{332}&1797&\textit{326}&2080&3.2$\times10^{-11}$\\
{3.77}&\textit{331}&{2760}&\textit{358}&1595&\textit{36}2&1513&\textit{36}0&1544&\textit{356}&1692&\textit{346}&1958&2.7$\times10^{-11}$\\
\rowcolor[HTML]{EFEFEF}[\tabcolsep] 
{4.78}&\textit{346}&{2492}&\textit{383}&1480&\textit{386}&1412&\textit{386}&1414&\textit{374}&1647&\textit{363}&1870&2.4$\times10^{-11}$\\
{5.90}&\textit{365}&{2265}&\textit{413}&1379&\textit{416}&1329&\textit{418}&1301&\textit{395}&1625&\textit{384}&1782&2.0$\times10^{-11}$\\
\end{tabular}
\caption{Measured temperature and corresponding thermal conductivity for different isotope ditributions}
\label{reftable}
\end{table*}

\begin{figure*}[htbp]

\includegraphics[width=.99\columnwidth]{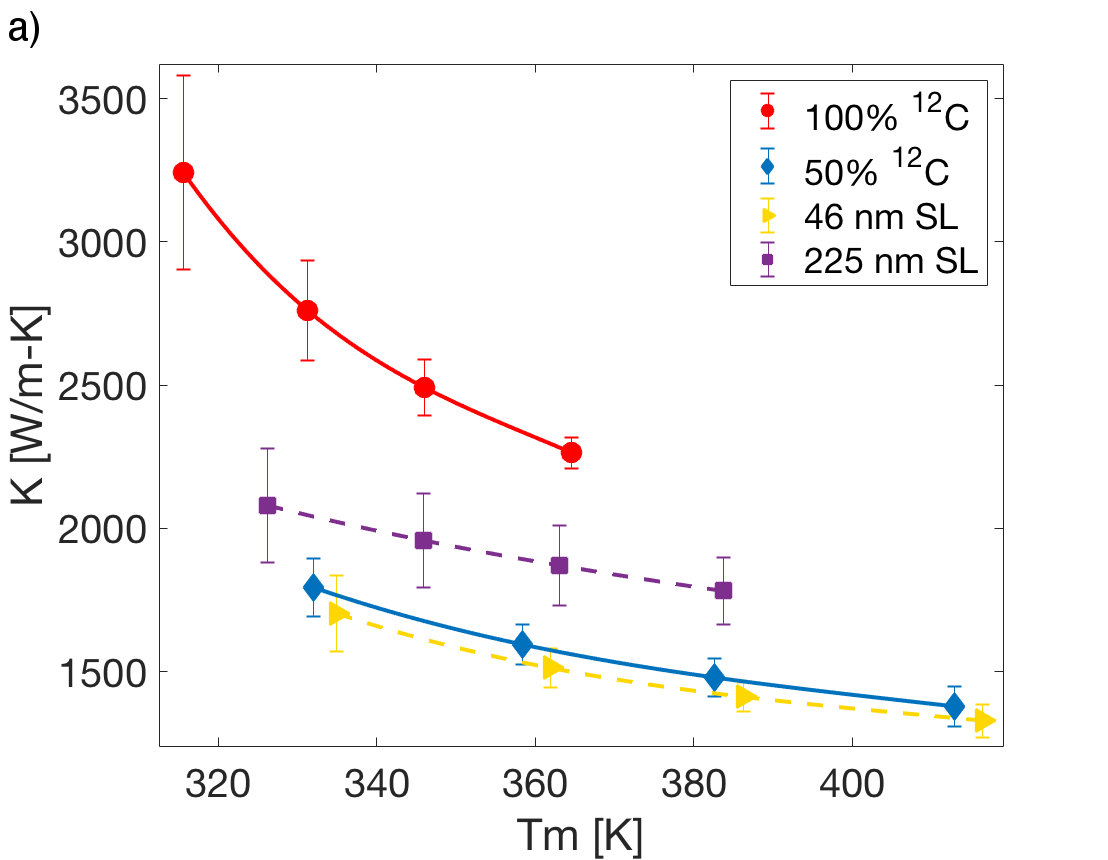}
\includegraphics[width=.99\columnwidth]{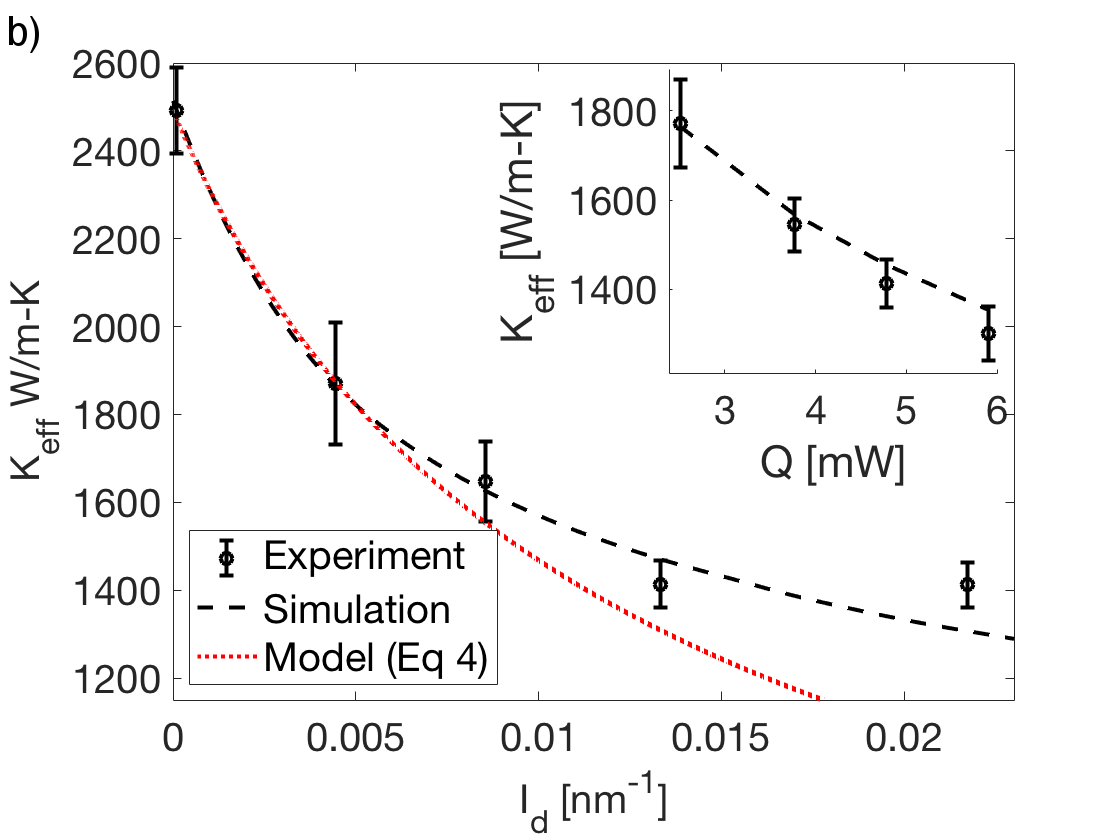}
\caption{a) Temperature dependence of the thermal conductivity of suspended graphene sheet with various \ctw/\cth isotope concentrations and distributions. (75 and 117 nm superlattices are omitted for clarity) b) Effective thermal conductivity as a function of interface density. Data points represent a single fixed heating power (4.78 mW). The dashed line shows the value of $K_{eff}$ determined from simulation setting $K_g = 2492$ W/m-K, and $R_{int}=2.4\times 10^{-11}\,\mathrm{m^2 K/W}$. The dotted line corresponds to equation \eqref{simplemodel} with the same parameters. The inset shows the dependence on heating power $Q$ for $L_s=117$ nm compared to the simulation (assuming a temperature independent $R_{int}$.}
\label{thermcond}
\end{figure*}

We note that a large part the quoted uncertainty on the thermal conductivity is due to the experimental uncertainty inherent to the Raman opto-thermal technique which is sensitive to the laser power output, graphene optical absorption and laser spot size and is dependent on the assumptions made in extracting the thermal conductivity from the measured spectrum. Experimental values of thermal conductivity of suspended graphene  measured by opto-thermal techniques cover a range from approx. 600 - 5000 W/m-K \cite{nika17,li17}, which reflects both sample quality variance and measurement technique differences in various experimental setups. Li et al. developed a technique to extract thermal conductivity independent of absorbed power by modifying laser spot size and find a value of thermal conductivity approx 1500 W/m-K \cite{li17} for single layer graphene. Whereas the uncertainty in the relative values of $K$,  determined from the std. error, are considerably smaller. In figure \ref{thermcond}b the error bars show the standard error and reflect the relative error between data points.

In the case of the superlattice samples the assumption of isotropic heat conduction is no longer valid and we should not expect a uniform value of $K$. In fact the thermal resistance may vary significantly when measured perpendicular and parallel directions relative to the mass periodicity. To first order we expect the effective thermal resistance for a distance $R$ in direction $\theta$ to be given by $R_{eff}(\theta)=R_g+\cos(\theta)\frac{2R}{L_s}R_{int}$, where $R_g$ is the thermal resistance of pristine graphene and $R_{int}$ the interfacial resistance (the average between the \ctw-\cth and \cth-\ctw interface). Here $\theta=0$ is the direction perpendicular to the \ctw-\cth interface. To obtain an estimate for our circular geometry, we can average the thermal conductivity over $\theta$ to obtain 
\begin{equation}
%R_{eff}\simeq R_g+\frac{4R}{\pi L_s}R_{int},
K_{eff}\simeq\frac{2L_s}{\pi R_{int}}\frac{\arctan\sqrt{\frac{a - 1}{a +1}}}{\sqrt{a^2-1}}
\label{simplemodel}
\end{equation}
where $R$ is the radius of the suspended graphene, $a=L_s/(2R_{int}K_g)$, and $R_g=RK_g^{-1}$. We expect this to be a good approximation for large periodicities ($L_s\gg r_0)$, where $r_0$ is the laser spot size. 

To obtain a more detailed picture taking into account the finite laser spot size and the non-uniform heat flow, we solve the inhomogoneous heat equation:

\begin{equation}
-\nabla \cdot (K(x,y) \nabla T) = \dot{q},
\end{equation}
where $\dot{q}$ is the volumetric heat source. The temperature is evaluated numerically using the relaxation method on a rectangular grid with spacing $h=1$ nm. Setting the initial temperature of the system at $T=293$ K and holding the boundary temperature fixed, the interior grid points are determined iteratively by: 

\begin{equation}
T_i^*=\sum_{\langle i j \rangle} \frac{K_jT_j}{\overline{K}_i} +h^2\frac{\dot{q}_i}{K_i}
\end{equation}

Where we have $\overline{K}_i=\sum_{\langle i j \rangle}K_{j}$ is the average thermal conductivity of the four nearest neighbours to i. 

\begin{figure}[htb]
\begin{center}
\includegraphics[width=.99\columnwidth]{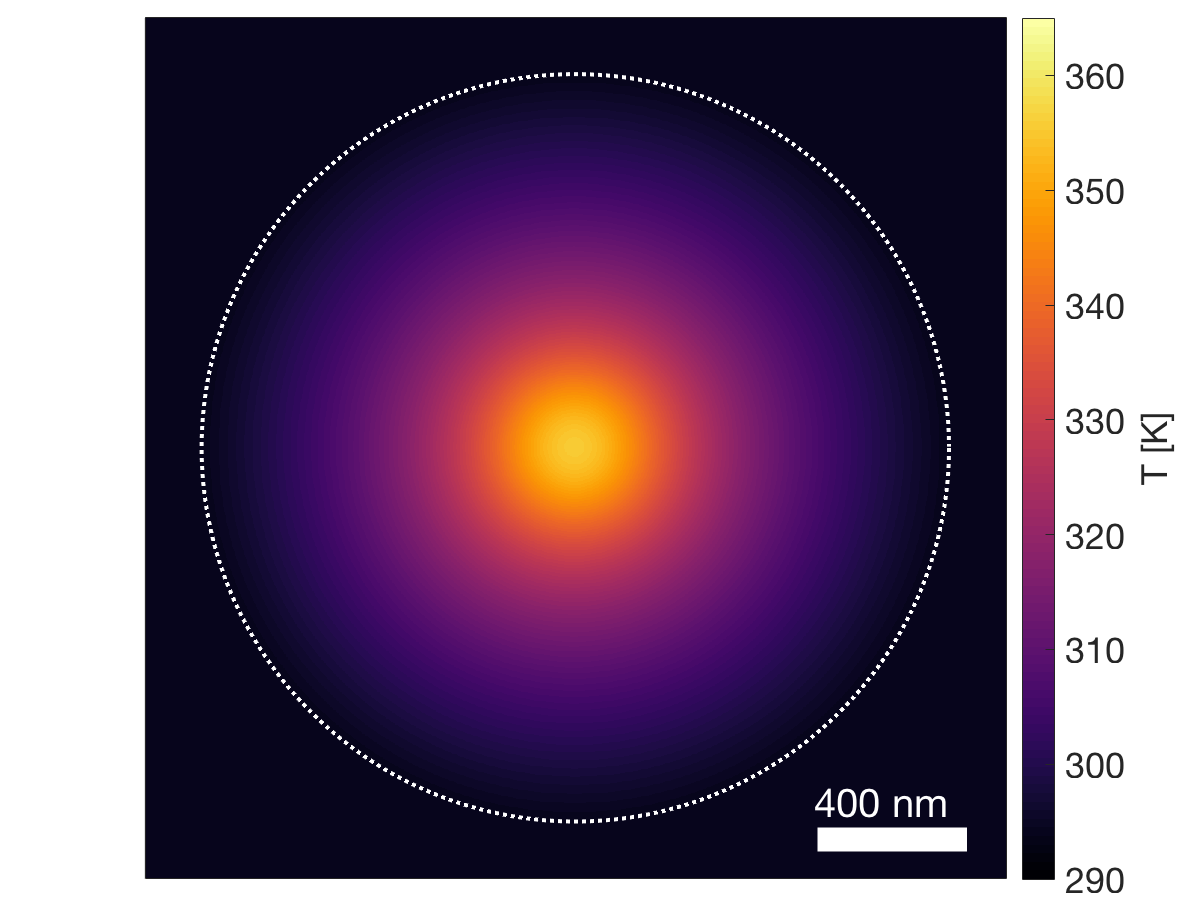}
\includegraphics[width=.99\columnwidth]{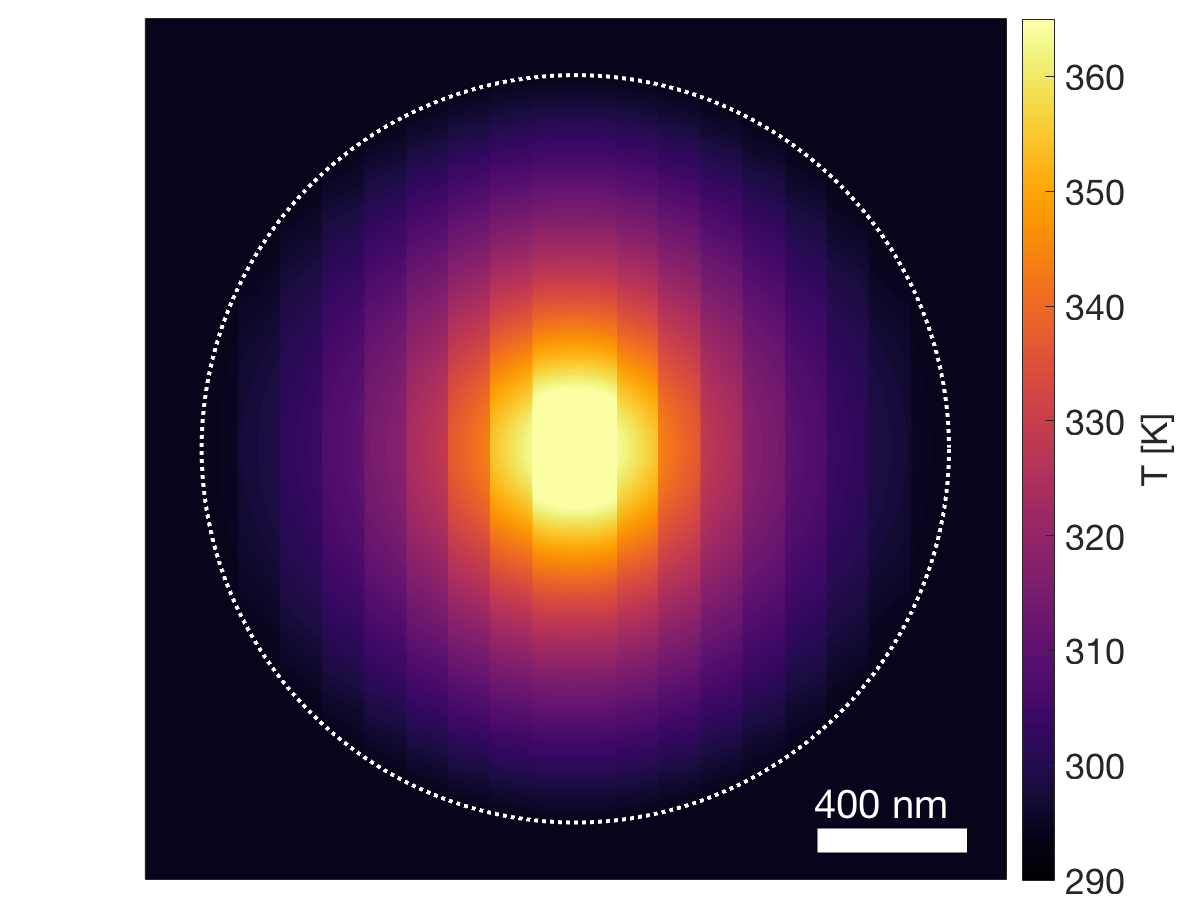}
\caption{Suspended graphene membrane temperature from simulation for \ctw graphene and 225 nm periodic superlattice with $Q=4.78$ mW heat source. Dashed line represents the boundary between the suspended and supported graphene. We find $T_m=345$ K and $T_m=363$ K for the \ctw and 225 nm SL resepctively. $K_g$ is taken to be the experimentally value 2492 W/m-K and $R_{int}$ is determined to be $2.4\times10^{-11}\,\mathrm{m^2 K/W}$}
\label{tmap}
\end{center}

\end{figure}

The periodicity dependence was modeled by considering a fixed graphene membrane thermal conductivity, $K_g$ with periodic interfaces represented by 1 nm strips with thermal resistivity $K_{int}=h/R_{int}$. We consider a circular suspended graphene membrane with radius $R=1\,\mu$m and thermal conductivity $K_g$ attached to a rectangular heatsink with thermal conductivity $K_{hs} \gg K_g$ for $R> 1 \mathrm{\mu m}$. In figure \ref{tmap} we show the temperature map across a homogeneous graphene membrane and a membrane with periodic interfaces with $L_s$ of 225 nm. Figure \ref{thermcond}a shows the effective thermal conductivity $K_{eff}$ given by equation \ref{k}, where the heat source $Q$ and measured temperature $T_m$ are both given by a Gaussian beam profile with diameter $D=340$ nm. These are compared to the experimentally measured values of $K_{eff}$ for the equivalent fixed laser power. The thermal conductivity of the film is taken as the experimental value $K_g=2492$ W/m-K and the best fit is obtained for an interfacial thermal resistance of $R_{int}=(2.4\pm 0.7)\times10^{-11}\,\mathrm{m^2 K/W}$. The value of $R_{int}$ varies depending on laser power, as shown in table \ref{reftable}, from 2.0-3.2 $\times10^{-11}\,\mathrm{m^2 K/W}$ which is comparable to the values found by non-equilibrium molecular dynamics studies, which find an interfacial thermal resistance of a graphene \ctw/\cth  interface as $1.05\times10^{-11}\,\mathrm{m^2 K/W}$\cite{mu15} and $3.88\times10^{-11}\,\mathrm{m^2 K/W}$\cite{pei12}. For the full range of experimental data, independent of laser power, the best fit is calculated as $R_{int}=(2.5\pm 0.5)\times10^{-11}\,\mathrm{m^2 K/W}$ 
% We also have to consider that our $R_{int}$ is an average between the \ctw/\cth and \cth/\ctw interface. 

When comparing the simulation with the experimental results we find a good agreement as shown in figure \ref{thermcond}b. The initial slope of the simulation agrees well with the estimate in equation \eqref{simplemodel} for large periods (225 nm). For smaller periods there is a systematic deviation between the simulation and \eqref{simplemodel}, which is due to the finite size of the heating area (laser spot size), where the simulation is closer to the experimental dependence. However, for the smallest period (46 nm) the experimental conductivity is larger than the value obtained by the simulation. This could be due to the coherence effects mentioned earlier, but is more likely due to the increased mixing of \ctw in the \cth phase and vice versa. In fact, molecular dynamics studies  of the thermal conductivity of a graphene isotope SL suggest that thermal conductivity can be further reduced by substituting additional isotope atoms on top of the periodic structure \cite{gu18}. However in a SL there is a trade-off between decreasing bulk conductivity by isotope impurities and increasing interface thermal conductance by reducing the mass difference between alternating isotope layers. Therefore at small periods isotope doping could lead to the observed increase in thermal conductivity. The total thermal resistance increase, is therefore a combination of residual isotope doping and interfacial thermal resistance.

\section{Conclusions}
The thermal conductivity of various isotope distributions in graphene show a reduction in thermal conductivity for both homogeneous isotope mixtures and superlattices. For pure \ctw graphene we measure thermal conductivity as high as 3200 W/m-K. In the case of periodic superlattices, the thermal conductivity decreases with increased interface density. The isotope interfacial thermal resistance is found to be $(2.5\pm 0.5) \mathrm{\times 10^{-11} \,m^2 K/W}$. In a polar geometry this leads to an almost factor 2 reduction in the thermal conductivity, while across the interfaces this reduction is even larger. Hence, we may expect a similar reduction for large polycrystalline CVD grown graphene superlattices, where interface orientation is randomized. The observed reduction in thermal conductivity can lead to interesting applications for thermoelectric devices that need high electrical conductivities with low thermal conductivities.  

\section{Acknowledgments:} We thank Jesse Maassen for helpful discussionss. This work was supported by NSERC, FRQNT, RQMP, CPM and INTRIQ.

\bibliography{ref}
\end{document}